\documentclass[showpacs,twocolumn,pre]{revtex4}
\usepackage{amsmath}
\usepackage{latexsym}
\usepackage{epsfig}
\usepackage{makeidx}
\usepackage{amsfonts}

\begin{document}

\title{Subdiffusion, superdiffusion and chemotaxis}
\author{Sergei Fedotov }
\affiliation{School of Mathematics, The University of Manchester, Manchester M60 1QD, UK }

\begin{abstract}
We propose two nonlinear random walk models which are suitable for the
analysis of both chemotaxis and anomalous transport. We derive the balance
equations for the population density for the case when the transition rate
for a random walk depends on residence time, chemotactic substance and
population density. We introduce the \textit{anomalous chemotactic sensitivity} and
find \textit{anomalous aggregation} phenomenon. So we suggest a new explanation of
the well-known effect of \textit{chemotactic collapse.} We develop a non-Markovian
"velocity-jump" model and obtain the superdiffusive behavior of bacteria with power law "run" time.
\end{abstract}

\maketitle

Continuous time random walks (CTRW) have been widely used in many fields
including physics, chemistry and life sciences (see, for example, excellent
reviews \cite{MK1,MK2}). Many biological and physical transport processes
exhibit anomalous behavior for which walker mean-squared displacement
increases as a fractional power $\mu $ of time: $<x^{2}(t)>\sim t^{\mu }$
(subdiffusion: $\mu <1$; superdiffusion: $\mu >1$). The chemotaxis is a
directed migration of cells population toward a more favorable environment.
The microscopic theory of the movement of cells or organisms is also based
on the random walk theory (see, for example, \cite{Alt,OS1,Hillen,EO}).
Although chemotaxis has a long history and has been studied by researchers
for many decades, there is a lack of literature on connection between the
anomalous random walk and chemotaxis theory. We should mention recent
exception \cite{LH} where biased CTRW has been analyzed. One of the reasons
for this gap is that the chemotaxis is essentially nonhomogeneous in space
and time random process, while the standard anomalous CTRW model involves
the spatial and temporal invariance \cite{MK1,MK2}. The main purpose of this
paper is to set up the random walk models for both chemotaxis and anomalous
transport. These models can be also used in other biological applications of
CTRW and chemotaxis as anomalous search strategy \cite{Vis}, cancer cell
dichotomy \cite{FI}, subdiffusion in spiny dendrites \cite{FM},
embryogenesis and wound healing. Much of the recent literature on chemotaxis
has been concerned \ with movement of bacteria \textit{E. coli} involving
the runs and tumbles. The standard assumption in modeling is that "run" and
"tumble" time intervals are exponentially distributed \cite{Hillen}. However
it has been found experimentally \cite{Kor} that the distribution of "run"
time intervals might deviate significantly from exponential approximation.
It has a power law and leads to superdiffusive behavior of bacteria. One of
the purposes of this paper is to examine \ the effect of long-time tails on
bacteria movement in terms of "velocity-jump" model.

We start with "space-jump" random walk model in one space dimension. The
cell performs a random walk as follows: it waits for a random time at each
point in space before making a jump to another point. The most important
characteristic of this walk is the transition rate $\gamma $ for jumps at
point $x.$ The standard assumption in CTRW theory is that $\gamma $ depends
on the residence time (age) $\tau $. This is a time interval between two
successive jumps of the cell. The corresponding waiting time density $\phi
(t)$ is related to $\gamma (\tau )$\ as $\phi (t)=\gamma (t)\exp \left(
\int_{0}^{t}\gamma (\tau )d\tau \right) $ \cite{Cox}. In chemotaxis theory
the jump of cells occurs in response to a chemical signal \cite{OS1}.
Therefore the transition rate $\gamma $ should depend on chemotactic
substance (external signal) $S(x,t)$ and its spatial and temporal gradient.
It also depends on macroscopic population density $\rho (x,t)$. This
dependence describes the coupling of the cells density and chemotactic
substance and crowding effects. Thus
\begin{equation}
\gamma (\tau |x,t)=\gamma (\tau |S(x,t),\dot{S}(x,t),\rho (x,t),t).
\end{equation}%
We introduce the cell density $\xi (x,t,\tau )$ at position $x$ at time $t$
with the residence time $\tau .$ The main reason for introduction of the
structured density $\xi $ is to make a random walk Markovian. This idea has
been used in \cite{Alt,Cox,VR,YH,MFH}. The density $\xi $ obeys the balance
equation
\begin{equation}
\frac{\partial \xi }{\partial t}+\frac{\partial \xi }{\partial \tau }%
=-\gamma (\tau |S(x,t),\dot{S}(x,t),\rho (x,t),t)\xi .  \label{basic}
\end{equation}%
We use the initial condition $\xi (x,0,\tau )=\rho _{0}(x)\delta (\tau )$
for which the residence time of all cells at $t=0$ equals to $0;$ $\rho
_{0}(x)$ is the initial density of cells. It is clear that the residence
time $\tau $ varies from $0$ to $t$. \ The condition at $\tau =0$ can be
written as%
\begin{equation}
\xi (x,t,0)=\int_{\mathbb{R}}\int_{0}^{t}\gamma (\tau |x,t)\xi (x-z,t,\tau
)w(z|x-z,t)d\tau dz.  \label{bau1}
\end{equation}%
Here $w(z|x,t)$ is the dispersal kernel for jumps $z$ which also depends on
chemotactic substance and its gradient, density $\rho (x,t)$ and $t$
\begin{equation}
w(z|x,t)=w(z|S(x,t),\dot{S}(x,t),\rho (x,t),t).  \label{jump}
\end{equation}%
It is assumed that $w$ is independent from $\tau .$ On the left hand side of
(\ref{bau1}) we have a density of cells just arriving at point $x$ at time $%
t $ (zero residence time). On the right hand side of (\ref{bau1}) we have an
integration of the rate at which the cells with different age $\tau $
arriving at position $x$ at time $t$ from the different points $x-z.$ Our
purpose now is to derive the Master equation for the cell density%
\begin{equation}
\rho (x,t)=\int_{0}^{t}\xi (x,t,\tau )d\tau  \label{dens}
\end{equation}%
Using the method of characteristics, we find from (\ref{basic}) that%
\begin{equation}
\xi (x,t,\tau )=\xi (x,t-\tau ,0)e^{-\int_{t-\tau }^{t}\gamma (s-(t-\tau
)|x,s)ds}.  \label{s1}
\end{equation}%
Let us denote the density of cells just arriving at point $x$ at time $t$ by
$j(x,t)=\xi (x,t,0).$We substitute (\ref{s1}) into (\ref{bau1}) and take
into account the initial condition for $\xi .$ We get%
\begin{equation}
j\left( x,t\right) =\int_{\mathbb{R}}i\left( x-z,t\right) w\left(
z|x-z,t\right) dz,  \label{ba0}
\end{equation}%
where the $i(x,t)$ is the density of cells leaving the point $x$ exactly at
time $t:$
\begin{equation}
i(x,t)=\int_{0}^{t}j(x,u)\phi (x,t,u)du+\rho _{0}(x)\phi (x,t,0),
\label{ba00}
\end{equation}%
\begin{equation}
\phi (x,t,u)=-\frac{\partial \Psi (x,t,u)_{{}}}{\partial t}=\gamma
(t-u|x,t)e^{-\int_{u}^{t}\gamma (s-u|x,s)ds}  \label{waiting}
\end{equation}%
and $\Psi (x,t,u)$ is the probability that a cell is trapped at point $x$
from time $u$ to $t$ without executing a jump%
\begin{equation}
\Psi (x,t,u)=e^{-\int_{u}^{t}\gamma (s-u|x,s)ds}.  \label{su}
\end{equation}%
This is an extension of standard survival function for a nonlinear and
nonhomogeneous case when $\Psi $ depends on chemotactic substance $S(x,t)$
and population density $\rho (x,t)$. The balance equation for $\rho \left(
x,t\right) $ can be found by substitution of (\ref{s1}) into (\ref{dens})
\begin{equation}
\rho \left( x,t\right) =\int_{0}^{t}j\left( x,u\right) \Psi (x,t,u)du+\rho
_{0}\left( x\right) \Psi (x,t,0).  \label{ba1}
\end{equation}%
The system of balance equations (\ref{ba0}), (\ref{ba00}) and (\ref{ba1}) is
a nonlinear generalization of classical CTRW renewal equations \cite{MK1,MFH}
and CTRW models for inhomogeneous and nonlinear media \cite{Ch,Sa}. These
equations can serve as a starting point for the analysis of both chemotaxis
and anomalous transport for "space-jump" random walk model. \ If we
differentiate $\rho \left( x,t\right) $ in (\ref{ba1}) with respect to time,
we obtain the nonlinear Master equation%
\begin{equation}
\frac{\partial \rho }{\partial t}=\int_{\mathbb{R}}i\left( x-z,t\right)
w\left( z|x-z,t\right) dz-i\left( x,t\right) .  \label{Master}
\end{equation}%
Now we are in a position to analyze the chemotaxis and anomalous effects in
more detail. First we consider the case when a cell performs a random walk
in a stationary environment with the distribution of \ chemotactic substance
$S(x)$. In this case $\gamma (\tau |x,t)=\gamma _{1}(\tau |S(x))$. The
survival probability $\Psi $ in (\ref{su}) must be a function of $\tau =t-u$
and can be written as
\begin{equation}
\Psi (\tau |S(x))=e^{-\int_{0}^{\tau }\gamma _{1}(u|S(x))du}.  \label{su1}
\end{equation}%
Using the Laplace transform in (\ref{ba0}), (\ref{ba00}) and (\ref{ba1}), we
obtain
\begin{equation}
i\left( x,t\right) =\int_{0}^{t}K_{x}\left( t-\tau \right) \rho \left(
x,\tau \right) d\tau ,  \label{ba38}
\end{equation}%
where $K_{x}(t)$ is the memory kernel defined by its Laplace transform
\begin{equation}
\hat{K}_{x}\left( s\right) =\frac{\hat{\phi}\left( s|S(x)\right) }{\hat{\Psi}%
\left( s|S(x)\right) },  \label{L}
\end{equation}%
where $s$ is the Laplace variable. Substitution of\ (\ref{ba38}) into (\ref%
{Master}) gives a generalized Master equation $\partial \rho /\partial
t=L_{_{1}}\rho $ with the operator $L_{_{1}}$:%
\begin{eqnarray}
L_{_{1}}\rho &=&\int_{0}^{t}\int_{\mathbb{R}}K_{x-z}\left( t-\tau \right)
\rho \left( x-z,\tau \right) w\left( z|x-z,t\right) dzd\tau  \notag \\
&&-\int_{0}^{t}K_{x}\left( t-\tau \right) \rho \left( x,\tau \right) d\tau .
\label{L1}
\end{eqnarray}%
The case when the dispersal kernel $w(z|x,t)$ depends on chemotactic
substance $S$ has been considered by Langlands and Henry \cite{LH}. It has
been pointed out by Erban and Othmer that movement of bacteria in favorable
environment is determined by chemokinesis rather than chemotaxis. In most
cases the bacteria or cell "does not feel" a macroscopic gradient of $S$
\cite{EO}. That is why it is more important to study the dependence of
transition probability $\gamma $ on chemotactic substance $S$ . To
illustrate the general theory we use only a symmetrical dispersal kernel $%
w(z)$ as a function of $z$.

In a Markovian case, when $\gamma $ does not depend on the residence time
variable $\tau $, we have $\Psi (x,t)=e^{-\gamma _{1}(S(x))t}$ and $\hat{K}%
_{x}\left( s\right) =\gamma _{1}(S(x)).$ Under the diffusion approximation,
the Master equation (\ref{L1}) takes the form
\begin{equation}
\frac{\partial \rho }{\partial t}=\frac{\sigma ^{2}}{2}\frac{\partial ^{2}}{%
\partial x^{2}}\left( \gamma _{1}(S(x))\rho \left( x,t\right) \right) ,
\label{di1}
\end{equation}%
where $\sigma ^{2}=\int_{\mathbb{R}}z^{2}w(z)dz$. It is well known \cite{OS1}
that this equation can be rewritten as $\partial \rho /\partial t+\partial
J/\partial x=0$ with the flux of cells
\begin{equation}
J=\chi \frac{\partial S}{\partial x}\rho -\frac{\sigma ^{2}\gamma _{1}(S(x))%
}{2}\frac{\partial \rho }{\partial x},\qquad \chi \left( S(x)\right) =-\frac{%
\sigma ^{2}}{2}\frac{\partial \gamma _{1}}{\partial S},
\end{equation}%
where $\chi $ is the chemotactic sensitivity. When the derivative $\partial
\gamma _{1}/\partial S$ is negative, the advection (taxis) is in the
direction of increase in chemotactic substance. In general it follows from (%
\ref{L1}) that cells flux is not local in time
\begin{eqnarray}
J &=&-\frac{\sigma ^{2}}{2}\frac{\partial S}{\partial x}\int_{0}^{t}\frac{%
\partial K_{x}\left( t-\tau \right) }{\partial S}\rho \left( x,\tau \right)
d\tau  \notag \\
&&-\frac{\sigma ^{2}}{2}\int_{0}^{t}K_{x}\left( t-\tau \right) \frac{%
\partial \rho \left( x,\tau \right) }{\partial x}d\tau .
\end{eqnarray}%
Instead of $\chi $ we have a chemotaxis memory kernel $\partial K_{x}\left(
t\right) /\partial S$. Note that the memory kernel for the chemotaxis flux
is different form the memory kernel for diffusion term (compare to \cite{LH}%
).

Let us consider the anomalous case when the waiting time PDF is
heavy-tailed, such that the corresponding mean time is infinite. We assume
that the longer cell survives at point $x$, the smaller the transition
probability from $x$ becomes. The rate $\gamma (\tau )$ is a monotonically
decreasing function of residence time $\tau .$ For example, if $\gamma (\tau
)=\mu \left( S(x)\right) /(\beta +\tau )$, it follows from (\ref{su1}) that
the survival function has a power-law dependence
\begin{equation*}
\Psi (\tau |S(x))=\left( \frac{\beta }{\beta +\tau }\right) ^{\mu \left(
S(x)\right) },
\end{equation*}%
where $\beta $ is constant. Anomalous case corresponds to $\mu \left(
S(x)\right) <1$ \cite{MK1,MFH}, when $\hat{K}_{x}\left( s|S(x)\right)
=s^{1-\mu \left( S(x)\right) }\tau _{0}^{-\mu \left( S(x)\right) };$ $\tau
_{0}$ is a parameter with units of time. The anomalous cell flux is%
\begin{eqnarray}
J &=&-\frac{\sigma ^{2}}{2}\frac{\partial S}{\partial x}\frac{\partial }{%
\partial S}\frac{1}{\tau _{0}^{\mu \left( S(x)\right) }}\mathcal{D}%
_{t}^{1-\mu \left( S(x)\right) }\rho \left( x,t\right)  \notag \\
&&-\frac{\sigma ^{2}}{2\tau _{0}^{\mu \left( S(x)\right) }}\mathcal{D}%
_{t}^{1-\mu \left( S(x)\right) }\frac{\partial \rho \left( x,t\right) }{%
\partial x},  \label{an}
\end{eqnarray}%
where the Riemann-Liouville fractional derivative $\mathcal{D}_{t}^{1-\mu
\left( S(x)\right) }$ is defined \cite{MK1,MFH} as%
\begin{equation}
\mathcal{D}_{t}^{1-\mu \left( S(x)\right) }\rho \left( x,t\right) =\frac{1}{%
\Gamma (\mu \left( S(x)\right) )}\frac{\partial }{\partial t}\int_{0}^{t}%
\frac{\rho \left( x,u\right) du}{(t-u)^{1-\mu \left( S(x)\right) }}.
\end{equation}%
It should be noted that the fractional time derivative of variable order $%
\mu (x)$ has been considered in \cite{Ch}. When $\mu =const$, we have a
classical subdiffusion transport equation for which the mean squared
displacement of cell increases with time as $t^{\mu }$ with $\mu <1.$

Let us consider the aggregation phenomenon \cite{OS1}. In a Markovian case,
in a finite domain with zero flux of cells on the boundary, there exists a
stationary non-uniform solution of (\ref{di1}) \cite{OS1}. The aggregation
of cells is due to the fact that mean waiting time $\gamma _{1}^{-1}(S(x)$
is decreasing function of the chemotactic substance $S$. In an anomalous
case, the system is not ergodic and there is no stationary distribution.
However, one can introduce the\textit{\ anomalous chemotactic sensitivity}
as a derivative of anomalous exponent: $\chi _{\mu }=\mu ^{\prime }\left(
S(x)\right) $. When $\chi _{\mu }<0,$ the cells will tend to aggregate where
the exponent $\mu $ is small. The anomalous flux (\ref{an}) leads to $\rho
\left( x,t\right) \rightarrow $ $\delta (x-x_{\min })$ as $t\rightarrow
\infty .$ Here $x_{\min }$ is the point where the anomalous exponent $\mu
\left( S(x)\right) $ has a minimum. It means that all cells aggregate into a
tiny region of space forming high density system at the point $x=x_{\min
\text{. }}$This phenomenon can be referred to as \textit{anomalous aggregation}%
. Similar results have been obtained in \cite{CFM} for a simple two-state
system. \ This effect is known in a literature as \textit{chemotactic
collapse} \cite{OS1}. Here we suggest an explanation of this effect which is
different from the classical one based on Keller-Segel equations. To prevent
the occurrence of delta-distribution we need to take into account the
crowding effect. In what follows, we consider this effect by assuming that
the transition rate depends on both the residence time $\tau $ and the
population density $\rho $.

If the transition rate $\gamma $ is independent of residence time $\tau ,$
then the system is Markovian. We assume that $\gamma $ depends on the time $%
t $ and the density $\rho \left( x,t\right) $ or non-stationary chemotactic
substance $S(x,t),$ that is $\gamma (\tau |x,t)=\gamma _{2}(\rho (x,t),t)$.
Then we obtain $i(x,t)=\gamma _{2}(\rho (x,t),t)\rho (x,t).$ The nonlinear
evolution equation for $\rho $ is $\partial \rho /\partial t=L_{_{2}}\rho ,$
where the operator $L_{_{2}}$ is defined as
\begin{eqnarray}
L_{2}\rho &=&\int_{\mathbb{R}}\gamma _{2}(\rho (x-z,t),t)\rho \left(
x-z,t\right) w\left( z|x-z,t\right) dz  \notag \\
&&-\gamma _{2}(\rho (x,t),t)\rho \left( x,t\right) .
\end{eqnarray}%
Now let us consider the case when the transition probability $\gamma (\tau
|x,t)$ depends both on the residence time $\tau $ and the density $\rho $ as
follows
\begin{equation}
\gamma (\tau |x,t)=\gamma _{1}(\tau |S(x))+\gamma _{2}(\rho (x,t),t).
\label{two}
\end{equation}%
From (\ref{ba0}), (\ref{ba00}) and (\ref{ba1}), after lengthy calculations,
we obtain
\begin{eqnarray}
i\left( x,t\right) &=&\int_{0}^{t}K_{x}\left( t-\tau \right) e^{-\int_{\tau
}^{t}\gamma _{2}(\rho (x,s),s)ds}\rho \left( x,\tau \right) d\tau  \notag \\
&&+\gamma _{2}(\rho (x,t),t).  \label{i2}
\end{eqnarray}%
It turns out that the nonlocal term in (\ref{i2}) involves the exponential
factor with $\gamma _{2}(\rho (x,t),t)$. Although $\gamma _{1}$ and $\gamma
_{2}$ are separable (see (\ref{two})), the corresponding terms in (\ref{i2})
are not separable. This is a non-Markovian memory effect. The generalized
Master equation is $\partial \rho /\partial t=L\rho ,$ where
\begin{eqnarray}
L\rho &=&\int_{0}^{t}\int_{\mathbb{R}}K_{x-z}\left( t-\tau \right) \rho
\left( x-z,\tau \right)  \label{main} \\
&&\times e^{-\int_{\tau }^{t}\gamma _{2}(\rho (x-z,s),s)ds}\times w\left(
z|x-z,t\right) dzd\tau -  \notag \\
&&\int_{0}^{t}K_{x}\left( t-\tau \right) \rho \left( x,\tau \right)
e^{-\int_{\tau }^{t}\gamma _{2}(\rho (x,s),s)ds}d\tau +L_{_{2}}\rho .  \notag
\end{eqnarray}%
It follows from here that $L\rho \neq L_{_{1}}\rho +L_{_{2}}\rho $ \ despite
the fact\ that\ $\gamma =\gamma _{1}+\gamma _{2}$. Similar phenomenon
related to chemical reactions has been discussed in \cite{YH,MFH,F}. The
exponential factor with $\gamma _{2}$ in (\ref{main}) prevents an \textit{%
anomalous aggregation} effect in a long-time limit.

Let us consider now $1$-D non-Markovian "velocity-jump" model for bacteria
movement. The purpose is to get the superdiffusive behavior \cite{Kor}. The
bacteria moves to the right with the velocity $v_{+}$ and reverses the
direction with the rate $\gamma _{+}.$ When the bacteria moves to the left
with the velocity $v_{-}$, the turning rate is $\gamma _{-}.$ In general,
the turning rate depends on run time $\tau ,$ on chemotactic substance $S$
and macroscopic population density $\rho (x,t):$ $\gamma _{\pm }(\tau |x,t)=$
$\ \gamma _{\pm }(\tau |S(x,t),\dot{S}(x,t),\rho (x,t),t).$ Let $\xi
_{+}(x,\tau ,t)$ be the density of bacteria moving with velocity $v_{+}$
with run time $\tau $. The corresponding density of organisms moving with
the velocity $v_{-}$ is $\xi _{-}(x,\tau ,t)$. Integration of $\xi _{\pm
}(x,t,\tau )$ over the run time variable $\tau $ gives the mean densities $%
\rho _{\pm }(x,t)=\int_{0}^{t}\xi _{\pm }(x,t,\tau )d\tau .$ The system of
equations for $\xi _{\pm }(x,\tau ,t)$ suggested by Alt \cite{Alt} are
\begin{equation}
\frac{\partial \xi _{\pm }}{\partial t}\pm v_{\pm }\frac{\partial \xi _{\pm }%
}{\partial x}+\frac{\partial \xi _{\pm }}{\partial \tau }=-\gamma _{\pm
}\left( \tau |x,t\right) \xi _{\pm }.  \label{dif1}
\end{equation}%
Initial conditions are $\xi _{\pm }(x,0,\tau )=\rho _{\pm }^{0}(x)\delta
(\tau )$, where $\rho _{\pm }^{0}(x)$ are the initial densities. Boundary
conditions at $\tau =0:$%
\begin{equation}
\xi _{\pm }(x,t,0)=\int_{0}^{t}\gamma _{\mp }(\tau |S(x,t),\rho (x,t),t)\xi
_{\mp }(x,t,\tau )d\tau .  \label{ba}
\end{equation}%
By using method of characteristics we solve (\ref{dif1}) and from (\ref{ba})
after lengthy manipulations we find the nonlinear system of equations for $%
\rho _{\pm }\left( x,t\right) $ and $j_{\pm }\left( x,t\right) =\xi _{\pm
}(x,t,0):$
\begin{eqnarray}
\rho _{\pm }\left( x,t\right) &=&\int_{0}^{t}j_{\pm }\left( x\mp v_{\pm
}(t-u),u\right) \Psi _{\pm }(x,t,u)du  \notag \\
&&+\rho _{\pm }^{0}\left( x\mp v_{\pm }t\right) \Psi _{\pm }(x,t,0),
\label{basic1}
\end{eqnarray}%
\begin{eqnarray}
j_{\pm }\left( x,t\right) &=&-\int_{0}^{t}j_{\mp }(x\pm v_{\mp }(t-u),u)\dot{%
\Psi}_{\mp }(x,t,u)du  \notag \\
&&-\rho _{\mp }^{0}\left( x\pm v_{\mp }t\right) \dot{\Psi}_{\mp }(x,t,0).
\label{basic2}
\end{eqnarray}%
Here we introduce the generalized survival function%
\begin{equation}
\Psi _{\pm }(x,t,u)=e^{-\int_{u}^{t}\gamma _{\pm }(s-u|x\mp v_{\pm
}(t-s),s)ds}
\end{equation}%
and its full derivative $\dot{\Psi}_{\pm }=$ $\frac{\partial \Psi _{\pm }}{%
\partial t}\pm v_{\pm }\frac{\partial \Psi _{\pm }}{\partial x}=$ $\ -\gamma
_{\pm }\left( \tau |S(x,t),\dot{S}(x,t),\rho (x,t),t\right) \Psi _{\pm }.$
If we differentiate $\rho _{\pm }\left( x,t\right) $ with respect to time,
we obtain the system of nonlinear equations
\begin{equation}
\frac{\partial \rho _{\pm }}{\partial t}\pm v_{\pm }\frac{\partial \rho
_{\pm }}{\partial x}=j_{\pm }\left( x,t\right) -j_{\mp }\left( x,t\right) .
\label{mean1}
\end{equation}%
If the switching rates $\gamma _{+}$ and $\gamma _{-}$ are independent of
run time $\tau $, then $j_{\pm }\left( x,t\right) =\gamma _{\mp }(\tau
|S(x,t),\dot{S}(x,t),\rho (x,t),t)\rho _{\mp }\left( x,t\right) .$ This
hyperbolic model has been studied by Hillen et al \cite{Hillen}.

Let us illustrate the general theory by considering the case when $v_{\pm }=v
$ and the run time PDF $\psi _{\pm }(\tau )=\gamma _{\pm }(\tau )\exp \left(
\int_{0}^{\tau }\gamma _{\pm }(\tau )d\tau \right) $ behaves like
\begin{equation}
\psi _{\pm }(\tau )\sim \left( \frac{\tau _{0}}{\tau }\right) ^{1+\mu
},\qquad \mu <1,\qquad \tau \rightarrow \infty .  \label{power}
\end{equation}
The mean waiting time $<\tau _{\pm }>=\int_{0}^{\infty }\tau \psi _{\pm
}(\tau )d\tau $ is infinite. The experimental evidence of a power-law
distribution like (\ref{power}) has been reported in \cite{Kor}. We assume
that $v_{+}=v_{-}=v$ and all bacteria run in a positive direction initially.
First we find the Laplace transform of $\left\langle x(t)\right\rangle :$ $%
\left\langle x(s)\right\rangle =-i\left( \frac{d\rho (k,s)}{dk}\right)
_{k=0},$ where $\rho (k,s)=\rho _{+}(k,s)+\rho _{-}(k,s)$ is the
Fourier-Laplace transform of the total bacteria density of particles $\rho
=\rho _{+}+\rho _{-}.$ In the limit $s\rightarrow 0,$ we find from (\ref%
{basic1}) and (\ref{basic2}) very unusual result that $\left\langle
x(s)\right\rangle \sim v\tau _{0}^{\mu }s^{-2+\mu }.$ It means that the
average position of bacteria is not zero as it should be in Markovian case!
It fact $\left\langle x(t)\right\rangle \sim v\tau _{0}^{\mu }t^{1-\mu }$ as
$t\rightarrow \infty $. The spreading is slower than a ballistic motion ($%
x(t)=vt$) and faster than diffusion for $0<\mu <0.5$ (superdiffusion).

In summary, we introduce two nonlinear CTRW models which are suitable for
the analysis of both chemotaxis and anomalous transport. We consider the
case when the transition rate for a random walk depends not only on
residence time, but also on chemotactic substance, its derivative and
macroscopic population density. We manage to derive the balance equations
for the population density and corresponding nonlinear Master equations. We
introduce the concept of \textit{anomalous chemotactic sensitivity} as a
derivative of anomalous exponent with respect to chemotaxis substance. We
find the effect of \textit{anomalous aggregation} when all bacteria tend to
aggregate at the point where power-law exponent has a minimum. So we suggest
a new explanation of \textit{chemotactic collapse } which is different from
the classical one based on Keller-Segel equations. Motivated by experiment
on run and tumble chemotaxis \cite{Kor}, we set up non-Markovian
"velocity-jump" model and obtain the superdiffusive behavior of bacteria
with power law "run" time.

\end{document}